\documentclass[%
aip,
jcp,
amsmath,amssymb,
reprint,%
groupedaddress
]{revtex4-1}

\usepackage{amssymb}
\newcommand\dif{\mathop{}\!\mathrm{d}}
\renewcommand{\vec}[1]{\mathbf{#1}}
\newcommand{\pdagger}{{\phantom{\dagger}}}
\newcommand{\exP}{\mathrm{e}^}
\usepackage{siunitx}
\usepackage{bm}
\usepackage{dsfont}

\usepackage{graphicx}
\usepackage{dcolumn}
\usepackage{bm}

\usepackage[english,capitalise]{cleveref}
\Crefname{equation}{Eq.}{Eqs.}
\Crefname{figure}{Fig.}{Figs.}
\Crefname{tabular}{Tab.}{Tabs.}
\usepackage[utf8]{inputenc}
\usepackage[T1]{fontenc}
\usepackage{mathptmx}
\usepackage{etoolbox}
\newcommand{\UniFreiburg}{Institute of Physics, Albert-Ludwig University Freiburg, Hermann-Herder-Strasse 3, 79104 Freiburg, Germany}

\makeatletter
\def\@email#1#2{%
	\endgroup
	\patchcmd{\titleblock@produce}
	{\frontmatter@RRAPformat}
	{\frontmatter@RRAPformat{\produce@RRAP{*#1\href{mailto:#2}{#2}}}\frontmatter@RRAPformat}
	{}{}
}%
\makeatother
\begin{document}
	
	\preprint{AIP/123-QED}
	
	\title[Influence of Nonequilibrium Vibrational Dynamics on Spin Selectivity in Chiral Molecular Junctions]{Influence of Nonequilibrium Vibrational Dynamics on Spin Selectivity in Chiral Molecular Junctions}
	\author{R. Smorka}
	\email{rudolf.smorka@physik.uni-freiburg.de.}
	\author{S. L. Rudge}
	\author{M. Thoss}
	\affiliation{%
		\UniFreiburg
	}%
	
	\date{\today}
	
	\begin{abstract}
		\noindent We explore the role of molecular vibrations in the chirality-induced spin selectivity (CISS) effect in the context of charge transport through a molecular nanojunction. We employ a mixed quantum-classical approach that combines Ehrenfest dynamics for molecular vibrations with the hierarchical equations of motion method for the electronic degrees of freedom. This approach treats the molecular vibrations in a nonequilibrium manner, which is crucial for the dynamics of molecular nanojunctions. To explore the effect of vibrational dynamics on spin selectivity, we also introduce a new figure of merit, the displacement polarization, which quantifies the difference in vibrational displacements for opposing lead magnetizations. We analyze the dynamics of single trajectories, investigating how the spin selectivity depends on voltage and electronic-vibrational coupling. Furthermore, we investigate the dynamics and temperature dependence of ensemble-averaged observables. 
		We demonstrate that spin selectivity is correlated in time with the vibrational polarization, indicating that dynamics of the molecular vibrations is the driving force of CISS in this model within the Ehrenfest approach.    	
	\end{abstract}
	\maketitle
	
	\section{Introduction}
	The chirality-induced spin selectivity (CISS) effect is the observation that the transport of spin-polarized electrons through a chiral medium depends on the handedness of the medium. This effect has been measured in a wide variety of experimental setups, such as spin-dependent photoemission \cite{gohler2011spin,kettner2018chirality,mollers2022chirality},
	electron transfer reactions \cite{eckvahl2023direct,abendroth2019spin,bloom2024chemical}, reactions of molecules on ferromagnetic surfaces \cite{spilsbury2023enantiosensitive,ghosh2020effect,naaman2018chirality}, and electron transport processes \cite{dor2013chiral,naaman2019chiral,aragones2017measuring,ortuno2023chiral,safari2023spin}.
	Despite the abundance of experimental evidence, the exact mechanism causing CISS remains unclear. 
	While spin-orbit coupling (SOC) is known to play a crucial role \cite{dalum2019theory,evers2022theory,naaman2020chiral}, calculations suggest that, if SOC were the sole contributor, it would need to be on the order of eV \cite{evers2022theory,guo2012spin,huisman2023chirality,geyer2019chirality}. However, carbon-based molecules, which show considerable spin selectivity in experiments, have SOC values in the range of just a few meV, even after accounting for factors like the molecular geometry and atomic SOC \cite{ando2000spin,huertas2006spin,dalum2019theory,naaman2020chiral,shitade2020geometric,geyer2020effective,medina2015continuum}. 
	This indicates that additional mechanisms are important to explain the CISS effect. 
	
	As a result, extensive theoretical efforts have been made to better understand the mechanisms underlying CISS \cite{bloom2024chiral,evers2022theory,naaman2020chiral}. These have explored aspects such as how to define the figure of merit \cite{liu2023spin}, environmental and interface effects \cite{dubi2022spinterface,naskar2023chiral,ghosh2020effect}, symmetry considerations \cite{zollner2020insight,yang2020detecting}, and the role of SOC in both the molecule and electrodes \cite{geyer2019chirality,zoellner2020_influence,naskar2022common}. Other aspects like molecular orientation \cite{ghazaryan2020analytic}, inelastic scattering \cite{huisman2021ciss,zhang2020chiral,guo2012spin}, temperature dependence \cite{das2022temperature,fransson2023temperature}, electron correlations \cite{fransson2020vibrational,fransson2022chiral,fransson2023chiral,huisman2023chirality,fransson2019chirality,huisman2021ciss,fransson2021charge,zhang2020chiral,klein2023giant,barroso2022spin,oppenheim2021incoherent,wu2021electronic,wang2021spin,shitade2020geometric,volosniev2021interplay}, and nonequilibrium conditions \cite{dalum2019theory} have also been investigated. Early studies have emphasized the importance of decoherence in breaking time-reversibility \cite{guo2012spin,matityahu2016spin,huisman2021ciss} and circumventing reciprocity relations that would otherwise prohibit spin selectivity \cite{bardarson2008proof,utsumi2020spin}. This has led to the hypothesis that CISS may be a many-body phenomenon driven by electron-electron or electronic-vibrational interactions \cite{volosniev2021interplay}.
	
	Building on this hypothesis, recent theoretical work has explored how the coupling between transport electrons and molecular vibrations could contribute to CISS. Models involving polaron transport \cite{klein2023giant,barroso2022spin,zhang2020chiral}, geometric effects such as due to Berry curvature \cite{wu2021electronic}, exchange splitting  \cite{fransson2020vibrational}, or chiral phonons \cite{fransson2023chiral,ohe2024chirality} suggest that electronic-vibrational interactions might be a key mechanism for CISS. 
	In particular, a model proposing vibrationally assisted SOC has been considered \cite{fransson2020vibrational}. 
	However, previous studies have generally treated molecular vibrations as being in thermal equilibrium, even under finite bias conditions \cite{fransson2020vibrational,das2022temperature}. Given that CISS is often observed under high bias conditions where the molecular vibrations are often far from equilibrium \cite{mishra2020length,mondal2020long,singh2024spin,thoss2018perspective,schinabeck2016hierarchical,rudge2024nonadiabatic,wang2020nonthermal}, this assumption requires further investigation.
	
	To this end, in this work, we analyze the model proposed in Ref.~\cite{fransson2020vibrational} with a method capable of treating nonequilibrium molecular vibrations. Since the full quantum problem is, at this point, too numerically challenging to solve for realistic system sizes, we employ the mixed quantum-classical Ehrenfest approach. This method treats the electronic degrees of freedom quantum mechanically via the hierarchical equations of motion (HEOM) approach and the molecular vibrations classically, with the electronic influence on the vibrations appearing as a mean-field force. 
	
	Using this approach, we demonstrate that a finite spin selectivity is correlated with highly nonequilibrium vibrational dynamics, which is measured via a quantity we call the displacement polarization. This quantifies the average difference in the displacement of the molecular vibrations from their equilibrium positions for opposing lead magnetizations. 
	The paper is organized as follows. We introduce the model of a chiral molecular junction and the mixed quantum-classical Ehrenfest and HEOM approach in Sec.~\ref{sec:Model}. In Sec.~\ref{sec:Results}, we present the results and discussion. We begin our analysis with the effect of classical molecular vibrations for a single trajectory at a fixed voltage in Sec.~\ref{sec:Results_B_1} and for a varying voltage in Sec.~\ref{sec:Results_B_2}, where we show the origin of electronic forces and how molecular vibrations result in a finite spin selectivity. Ensemble averages of currents, spin selectivity and total vibrational bond displacement are presented and discussed in Sec.~\ref{sec:Results_C}. The paper is concluded in Sec.~\ref{sec:Conclusion}.
	
	\section{Model and Method}\label{sec:Model}
	
	In this section, we introduce both the model of a chiral molecular junction as well as the mixed quantum-classical approach we use to calculate the dynamics. Furthermore, in the last part of this section, we introduce the observables of interest for the CISS problem, which includes the standard figure of merit as well as a new parameter that helps us to characterize the polarization of the vibrational degrees of freedom. 
	
	\subsection{Model of a Chiral Nanojunction}
	
\begin{figure}[!h]
		\centering
		\includegraphics[width=\columnwidth]{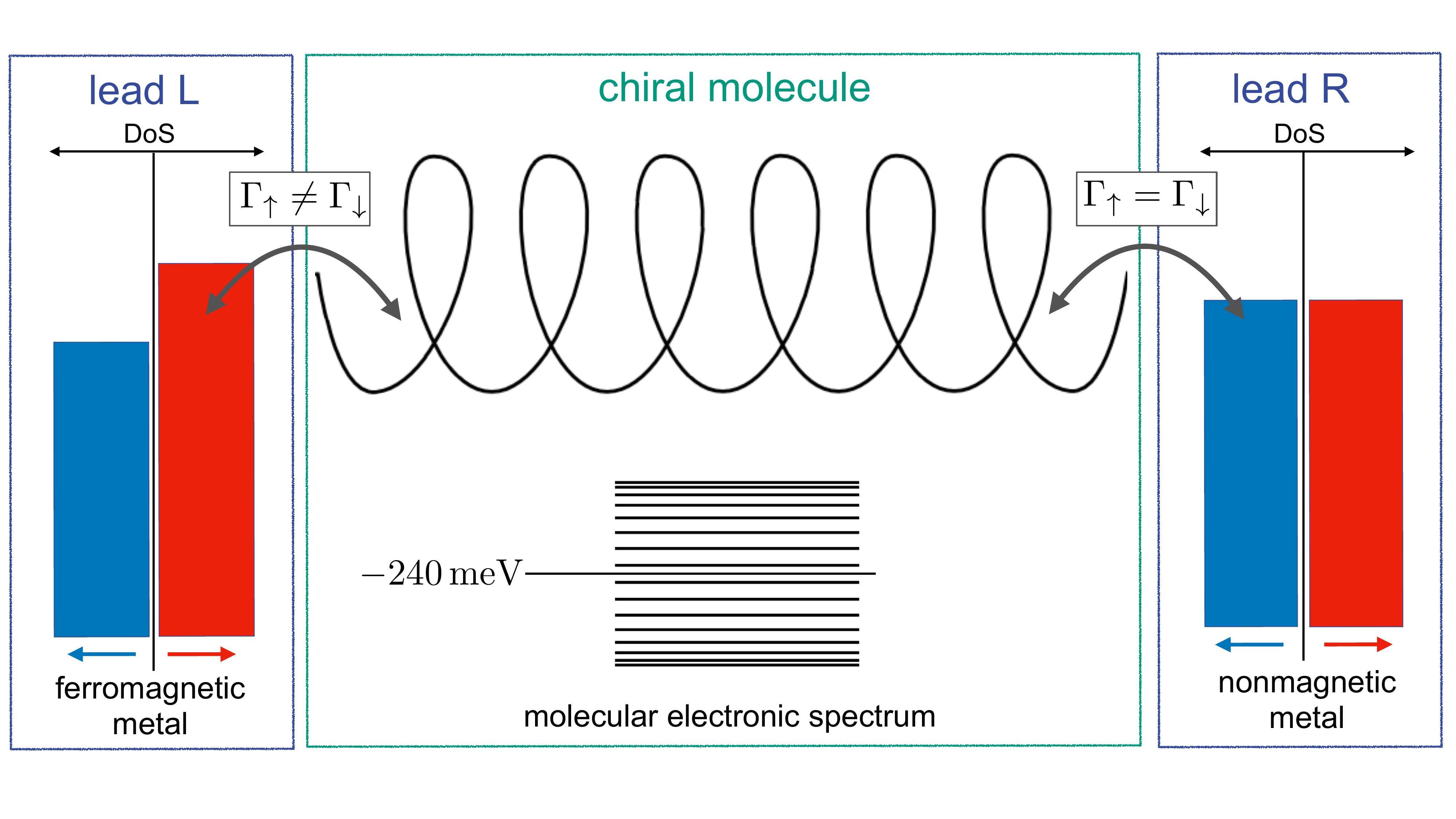}
		\caption{Schematic of the chiral molecular junction, consisting of a chiral molecule (center) coupled to a ferromagnetic lead $L$ (left) and a nonmagnetic lead $R$ (right).}\label{fig:Schematic}
	\end{figure}

	We consider a model based on that suggested in Ref.~\cite{fransson2020vibrational}, which describes a molecular junction consisting of a chiral molecule connected to two macroscopic leads. The setup is illustrated schematically in \cref{fig:Schematic}. 
	
	The Hamiltonian, $H$, of the overall system is composed of three parts,
	\begin{equation}
		\begin{split}
			H &= H_\mathrm{mol} +   H_\mathrm{leads}+H_\mathrm{mol-leads}.
		\end{split}\label{eq:4_Hamiltonian_Quantum_Open}
	\end{equation}
	The molecular part, $H_\mathrm{mol}$, is described by the model of a chiral molecule introduced in Ref.~\cite{fransson2020vibrational}, 
	\begin{align}
		H_\mathrm{mol} &= H_\mathrm{el}^{(0)}+H_\mathrm{vib}+H_\mathrm{el-vib},\label{eq:H_mol}\\
		H_\mathrm{el}^{(0)}	&=
		\begin{aligned}[t]
			&\sum_{j\sigma} \varepsilon_{0}^\pdagger d_{j\sigma}^\dagger d_{j\sigma}^\pdagger -  \sum_{j\sigma} t_0^\pdagger (d_{j\sigma}^\dagger d_{j+1,\sigma}^\pdagger + \mathrm{h.c.})+\\
			& \sum_{j\sigma\sigma'} \lambda_0^\pdagger(id_{j\sigma}^\dagger (\vec{v}_j^{(+)}\cdot \bm{\sigma})_{\sigma\sigma'}^\pdagger d_{j+2,\sigma'}^\pdagger + \mathrm{h.c.}),
		\end{aligned}
		\label{eq:H_el}\\[\jot]
		H_\mathrm{vib} &= \sum_{j}\frac{\omega_j}{2}\left(X_j^2 + P_j^2 \right),\label{eq:H_vib}\\	
		H_\mathrm{el-vib}	&=
		\begin{aligned}[t]
			&\sum_{j\sigma} \frac{t_1}{\sqrt{2}}^\pdagger X_j^\pdagger (d_{j\sigma}^\dagger d_{j+1,\sigma}^\pdagger + \mathrm{h.c.}) + \\
			& \sum_{j\sigma\sigma'} \frac{\lambda_1}{\sqrt{2}}^\pdagger X_j^\pdagger (id_{j\sigma}^\dagger (\vec{v}_j^{(+)}\cdot \bm{\sigma})_{\sigma\sigma'}^\pdagger d_{j+2,\sigma'}^\pdagger + \mathrm{h.c.}).
		\end{aligned} \label{eq:H_el-vib}
	\end{align}
	The chiral molecule is modeled as a collection of electronic sites labeled by $j$ with spin $\sigma \in \{\uparrow,\downarrow\}$. The purely electronic part of the molecule, $H_\mathrm{el}^{(0)}$, is described within a tight-binding-like framework. 
	Here, $d_{j\sigma}^\dagger$ and $d_{j\sigma}^\pdagger$ denote fermionic creation and annihilation operators at site $j$ with spin $\sigma$. 
	The sites can be thought of as the constituents of the chiral molecule, such as monomers in the case of a chiral polymer, contributing with a single molecular orbital per constituent. For simplicity, it is assumed that each site has the same energy, $\varepsilon_{0}$. The nearest-neighbor hopping amplitude and the Rashba-type SOC strength are given by $t_{0}$ and $\lambda_{0}$, respectively  \cite{fransson2019chirality,fransson2020vibrational,varela2016effective}. 
	The vector $\vec{v}_j^{(s)}=\vec{d}_{j+s}\times \vec{d}_{j+2s}$ is defined by the geometry of the chiral molecule in terms of unit vectors $\vec{d}_{j+s}=(\vec{r}_j-\vec{r}_{j+s})/\vert \vec{r}_j - \vec{r}_{j+s}\vert$, where $\vec{r}_j =(a\cos\phi_j,a\sin\phi_j,h_j)$ are the position vectors to the sites of the helix with radius $a$ and pitch $h$. The helix consists of $N_\mathrm{laps}$ number of windings, $N_\mathrm{sites}$ sites per winding, and $N=N_\mathrm{laps}\times N_\mathrm{sites}$ total number of sites. Within the helix, therefore, the sites are also characterized by their polar angle, $\phi_j= 2\pi(j-1)N_\mathrm{laps}/(N-1)$, and height, $h_j = (j-1)h/(N-1)$ of site $j$. 
	
	The chiral molecule also contains inter-site vibrational modes, which are described by $H_\mathrm{vib}$ in \cref{eq:H_vib}. Here, $X_j$ and $P_j$ are the dimensionless displacement and momentum operators, respectively, and $\omega_j$ denotes the frequency of the vibrational mode coupling sites $j$ and $j+1$. Physically, the vibrational degree of freedom could describe the distance between the centers of mass of neighboring constituents of the chiral molecule \cite{conwell2000polarons,chakraborty2007charge}. 
	
	The vibrational modes couple to the electronic sites via $H_\mathrm{el-vib}$, which is given in \cref{eq:H_el-vib}. This term consists of two contributions. 
	The first is a spin-independent nearest-neighbor electronic coupling to the bond vibrational modes with a coupling strength denoted by $t_1$. This contribution to the electronic-vibrational coupling only is commonly referred to as the Su-Schrieffer-Heeger model \cite{su1979solitons,troisi2006charge}. 
	The second contribution is a linear spin-dependent coupling, similar in form to the SOC with coupling strength denoted by $\lambda_1$. 
	These contributions can be understood as resulting from a first-order expansion of the hopping and spin-orbit overlap integrals in terms of the displacement between sites \cite{su1979solitons,fransson2020vibrational,fransson2019chirality,das2022temperature}. 
	In the following analysis, we will also employ the definition of the electronic Hamiltonian containing the interaction part, 
	\begin{equation}
		H_\mathrm{el}(X)=H_\mathrm{el}^{(0)}+H_\mathrm{el-vib}(X).\label{eq:H_el_t}
	\end{equation}
	
	The chiral molecule is connected to two leads, which are modeled as noninteracting electrons, 
	\begin{align}
		H_\mathrm{leads} = \sum_{\ell \in \{L,R\}} \sum_{\alpha,\sigma} \varepsilon_{\ell \alpha \sigma}^\pdagger c_{ \ell \alpha \sigma}^\dagger c_{\ell \alpha \sigma}^\pdagger.
	\end{align}
	Here, the operators $c_{\ell \alpha \sigma}^\dagger$ and $c_{ \ell \alpha \sigma}^\pdagger$ create and annihilate an electron in state $\alpha$ of lead $\ell$ and with spin $\sigma$, respectively, with $\varepsilon_{\ell \alpha \sigma}^\pdagger$ being the corresponding energy. 
	
	The left lead couples linearly and in a spin-dependent fashion to the leftmost site of the chiral molecule, while the right lead couples to the corresponding rightmost site, 
	\begin{align}
		H_{\mathrm{mol}-\mathrm{leads}} = 
		&\sum_{\alpha, \sigma} \big(\gamma_{L \alpha \sigma}^\pdagger d_{1\sigma}^\dagger c_{L \alpha \sigma}^\pdagger + \gamma_{R \alpha \sigma}^\pdagger d_{N\sigma}^\dagger c_{R \alpha \sigma}^\pdagger+\mathrm{h.c.}\big),
	\end{align}
	where $\gamma_{\ell\alpha\sigma}$ are the spin-dependent coupling strengths. The leads are also characterized by the level-width functions \cite{haug2008quantum,ryndyk2016theory},
	\begin{equation}
		\begin{split}
			\Gamma_\ell(\varepsilon) = 2\pi\sum_{\alpha, \sigma} \vert \gamma_{\ell\alpha \sigma} \vert^2 \delta(\varepsilon-\varepsilon_{\ell\alpha\sigma}).
		\end{split}
	\end{equation} 
	In this work, we assume the wide-band limit, in which the level-width function is energy independent, $\Gamma_\ell =\mathrm{const}$. In order to inject electrons of a particular spin to the chiral molecule, we model the left lead as ferromagnetic while keeping the right lead nonmagnetic. This is done by introducing a spin-dependent level-width function for the left lead,
	\begin{equation}
		\Gamma_L^A= \Gamma_L(\mathds{1}+A\sigma^z),\label{eq:Gamma_A}
	\end{equation}
	with $A=\pm \frac{1}{2}$ and $\sigma^z$ the Pauli $z$ matrix, whereas the right lead is unpolarized, that is $\Gamma_R$ is diagonal in the spin subspace  \cite{fransson2020vibrational,fransson2022chiral}.  
	Both $\Gamma_L$ and $\Gamma_R$ are diagonal $2N\times 2N$ matrices with non-vanishing matrix elements $(\Gamma_L)_{1\sigma}$ and $(\Gamma_R)_{N\sigma}$ for both spins $\sigma$, and zeroes elsewhere.
	The asymmetry in the molecule-lead coupling ensures that electrons of the majority spin in the ferromagnetic lead are injected with a higher probability into the molecule than electrons with minority spin. Spin-flip processes between the leads and the molecule are excluded.
	
	\subsection{Ehrenfest-HEOM Method \label{sec:Method}}
	
	To calculate transport properties of the chiral molecular junction, we employ a mixed quantum-classical approach. The hierarchical equations of motion (HEOM) technique \cite{tanimura1989time,tanimura2006stochastic,welack2006influence,popescu2013treatment,popescu2016efficient,haertle2013decoherence,haertle2014formation,haertle2015transport,wilkins2015quantum,jin2008exact,li2012hierarchical,zheng2007time,zheng2012hierarchical,ye2016heom,tanimura2020numeric} is used to propagate the electronic degrees of freedom quantum mechanically, whereas the vibrational degrees of freedom are treated classically within the Ehrenfest approach \cite{stock2005classical,horsfield2004open,verdozzi2006classical,todorov2010nonconservative,subotnik2010augmented,todorov2014current,tully1998nonadiabatic,kirrander2020ehrenfest}. This mixed quantum-classical method \cite{tully1998mixed,stock2005classical,elze2012linear} was introduced in Ref.~\cite{erpenbeck2018current}, based on the derivation of the HEOM within the context of quantum transport \cite{haertle2013decoherence,jin2008exact,zheng2012hierarchical,croy2009propagation}. 
	The equations of motion given below can also be derived using the Keldysh formalism applied to the nonequilibrium Green's function technique. Therefore, the electronic contribution referred to as HEOM in this paper is also known as time-dependent nonequilibrium Green's functions 
	in the literature \cite{wingreen1993time,jauho1994time,ryndyk2009green,croy2009propagation,popescu2013treatment, rahman2018non, popescu2016efficient}. 
	This equivalence holds, however, only for non-interacting systems. 
	
	From a quantum perspective, the central object of interest is the reduced density matrix of the electronic part of the chiral molecule, $\rho_\mathrm{el}$, which is obtained by tracing out the degrees of freedom in the leads from the total density matrix of the nanojunction, $\rho$: $\rho_\mathrm{el}=\mathrm{tr}_\mathrm{leads}\{\rho\}$. Within the HEOM framework, for an initially uncorrelated density operator $\rho=\rho_\mathrm{el}\otimes \rho_\mathrm{leads}$, the time evolution of $\rho_\mathrm{el}$ is given by 
	\begin{equation}
		i\hbar \frac{\partial }{\partial t} \rho_\mathrm{el}(t) = [H_\mathrm{el}(X),\rho_\mathrm{el}(t)] + i \sum_\ell \left(\Pi^\dagger_\ell(t) + \Pi_\ell^\pdagger(t) \right). \label{eq:RhoOpenEOM}
	\end{equation}
	The first term in \cref{eq:RhoOpenEOM} describes the unitary time evolution due to the Hamiltonian of the chiral molecule, which acquires an implicit time dependence due to the presence of the vibrational coordinates, $X_j(t)$. The second term on the right-hand side of \cref{eq:RhoOpenEOM} describes the dissipative, nonunitary time evolution of the reduced density matrix due to coupling of the molecule to the leads, which takes the form of the so-called auxilliary density matrices (ADMs), $\Pi_\ell(t)$. 
	
	Since the electronic part of the molecular Hamiltonian is noninteracting, these ADMs can also be expressed in terms of time-dependent nonequilibrium Green's functions (NEGFs) \cite{smorka2022nonequilibrium,smorka2024dynamics,rahman2018non,leitherer2017simulation,popescu2016efficient,haug2008quantum}. Utilizing the Padé-decomposition of the leads' Fermi-Dirac distributions \cite{hu2010communication}, one obtains the form  
	\begin{equation}
		\Pi_\ell(t) = \frac{1}{4}(\mathds{1}-2\rho_\mathrm{el}(t))\Gamma_\ell+\sum_{p=1}^{N_p} \Pi_{\ell,p}(t).\label{eq:AuxiliaryRhoDef}
	\end{equation}
	In \cref{eq:AuxiliaryRhoDef}, $N_p$ is the number of Padé poles used in the decomposition and $\Pi_{\ell,p}(t)$ are the Padé-resolved ADMs of lead $\ell$, which follow the equation of motion~\cite{croy2009propagation,popescu2016efficient,leitherer2017simulation}
	\begin{equation}
		\begin{split}
			i\hbar \frac{\partial }{\partial t}\Pi_{\ell,p}(t) = &\frac{\eta_p}{\beta}\Gamma_\ell +  \left(H_\mathrm{el}(t)-\frac{i}{2}\Gamma-\chi_{\ell, p}^+\mathds{1}\right)\Pi_{\ell,p}(t).\label{eq:PadeAuxiliaryRhoEOM}
		\end{split}
	\end{equation}
	Here, $\Gamma=\sum_\ell\Gamma_\ell$ denotes the total level-width function, while $\eta_p$ are the coefficients of the Padé expansion. The corresponding Padé poles, $\xi_p$, enter into the frequencies, $\chi_{\ell,p}^+$, via $\chi_{\ell,p}^+ = \mu_\ell + i\xi_p\beta_\ell^{-1}$. The frequencies also depend on the chemical potentials, $\mu_\ell$, and inverse temperatures, $\beta_\ell = 1/(k_B T_\ell)$, of the leads. In this work, a DC bias voltage, $\Phi$, is applied as a symmetric chemical potential difference: $\mu_L=E_F + \frac{e\Phi}{2}$ and $\mu_R=E_F - \frac{e\Phi}{2}$, where $E_F = 0$ eV is the Fermi energy of the junction.
	The HEOM given in \cref{eq:RhoOpenEOM,,eq:AuxiliaryRhoDef,,eq:PadeAuxiliaryRhoEOM} are numerically exact as long as the Fermi-Dirac function is accurately represented by the Pad\'{e} decomposition. For this, a sufficient number of Pad\'{e} poles is required, which depends on the voltage bias and the temperature of the leads. In the following calculations, $N_p=15$ poles have been used. 
	
	We treat the dynamics of molecular vibrations within the Ehrenfest method.
	This approach has been applied to study nuclear dynamics in quantum transport in Refs.~\cite{verdozzi2006classical,metelmann2011adiabaticity}.
	Specifically, the Ehrenfest-HEOM method has been used in the context of transport through molecular junctions to investigate phenomena such as current-induced bond rupture
	\cite{erpenbeck2018current}, nonequilibrium transport through polymer chains \cite{zhang2024effects}, and relaxation dynamics and transport in spin valves \cite{smorka2022nonequilibrium,smorka2024dynamics,petrovic2018spin}.
	Additionally, this approach has been applied without incorporating feedback effects in Refs.~\cite{leitherer2017simulation,rahman2019dephasing,mejia2022coherent}.
	
	The time-evolution of the vibrational coordinates and momenta within the Ehrenfest approach is given by the classical equations of motion
	\begin{align}
		\dot{X}_j &= \mathrm{tr} \left\{\rho_\mathrm{el}(t) \frac{\partial H_\mathrm{mol}}{\partial P_j}\right\} = \omega_j P_j, \label{eq:EOM_Classical_1}\\
		\begin{split}
			\dot{P}_j &= -\mathrm{tr}\left\{\rho_\mathrm{el}(t) \frac{\partial H_\mathrm{mol}}{\partial X_j}\right\} \\
			&= -  \omega_jX_j - \mathrm{tr}\left\{\rho_\mathrm{el}(t)\frac{\partial H_\mathrm{el-vib}(X)}{\partial X_j}\right\}.\label{eq:EOM_Classical_2}
		\end{split}
	\end{align}
	The combined Ehrenfest-HEOM method thus consists of solving the coupled set of equations of motion \cref{eq:RhoOpenEOM,,eq:AuxiliaryRhoDef,,eq:PadeAuxiliaryRhoEOM} in the quantum sector and simultaneously the classical equations of motion \cref{eq:EOM_Classical_1,eq:EOM_Classical_2}.
	
	The equations are solved numerically as an initial value problem. The electronic equations of motion, \cref{eq:RhoOpenEOM,,eq:AuxiliaryRhoDef,,eq:PadeAuxiliaryRhoEOM}, are propagated via standard fourth-order Runge-Kutta with a fixed time step. The classical equations of motion, meanwhile, \cref{eq:EOM_Classical_1,eq:EOM_Classical_2}, are solved using a fourth order symplectic integrator \cite{forest1990fourth}. This hybrid integration scheme has a number of advantages. First, it conserves energy in an isolated system ($\Gamma_0=0$), contrary to a scheme employing lower-order symplectic integrators or non-symplectic integrators for the vibrational degrees of freedom. The second advantage is that such a symplectic scheme is numerically stable even for very long propagation times, which is particularly desirable for systems involving both high-frequency electronic dynamics on the femtosecond timescale and low-frequency vibrational modes with small couplings, which have a timescale in the picoseconds.
	
	\subsection{Observables} \label{sec:Method_C}
	
	While different definitions of the figure of merit for CISS exist for different experimental setups, in the context of electron transport, spin selectivity is often measured in terms of charge currents for different lead polarizations. At the interface between lead $\ell$ and the molecule, the charge current, $I_{\ell}(t)$, is defined as the expectation value of the time-derivative of the particle number operator $N_\ell = \sum_{\alpha,\sigma}c_{\ell\alpha\sigma}^\dagger c_{\ell\alpha\sigma}^\pdagger$ in lead $\ell$,
	\begin{align}
		I_\ell(t) \equiv -e\left\langle \frac{\dif N_\ell}{\dif t}\right\rangle &= -\frac{ie}{\hbar}\langle [H,N_\ell]\rangle,
	\end{align} 
	which can be obtained directly from the ADMs within the HEOM approach (\cref{eq:AuxiliaryRhoDef}) \cite{popescu2013treatment,croy2009propagation},
	\begin{align}
		I_\ell(t) &=	\frac{2e}{\hbar}\mathrm{Re}\, \mathrm{tr}\{\Pi_\ell(t)\}\label{eq:DefCurrent}.
	\end{align}
	Next, we denote by $I_{L \pm}(t)$ the charge current through the interface of the molecule and the left lead $L$ obtained in a setup with magnetization asymmetry $A=\pm\frac{1}{2}$ in the molecule-lead coupling $\Gamma_L^A$ (cf. \cref{eq:Gamma_A}). From this, we can define the spin-polarized net charge current of the junction,
	\begin{equation}
		I_\pm(t) = \frac{I_{L \pm}(t)-I_{R}(t)}{2}.\label{eq:I_p}
	\end{equation}
	The strength of the CISS effect is characterized by the so-called spin selectivity, $S$ \footnote{The spin selectivity $S$ is also called \emph{spin polarization} in the literature and is sometimes denoted by $P$.}.
	We define the spin selectivity figure of merit, $S(t)$, as the normalized difference of charge currents for opposing lead magnetizations,
	\begin{equation}
		S(t) = \frac{I_+(t)-I_-(t)}{I_+(t)+I_-(t)}.\label{eq:SpinSelectivity}
	\end{equation}
	This quantity requires the measurement of charge currents, which are the natural observables in transport experiments. The spin selectivity in \cref{eq:SpinSelectivity} has been related to the magnetoresistance within a linear response treatment \cite{mondal2016spin,naaman2015spintronics,liu2023spin}. Note, however, that our treatment here is not limited to the linear-response regime, and that $S(t)$ is time-dependent since $I(t)$ is time-dependent.
	
	To relate the spin selectivity dynamics with the vibrational dynamics of the molecule, we introduce an observable that quantifies the influence of the lead magnetization on the vibrational displacement, which we call the displacement polarization,
	\begin{equation}
		S_X(t) = \bar{X}_+(t)-\bar{X}_-(t).
	\end{equation}
	We define the displacement polarization as the difference in average displacements $\bar{X}_\pm$ for opposing lead magnetizations, where the average displacement, $\bar{X}(t)$, is defined as
	\begin{equation}
		\bar{X}(t) = \frac{1}{N-1}\sum_j X_j(t).\label{eq:X_avg}
	\end{equation} 
	Note, that for $N$ electronic sites, there are $N-1$ vibrational modes in the model considered here. The displacement polarization provides a clear and quantifiable measure of the influence of lead magnetization on the vibrational dynamics of the molecule. It also allows us to analyze symmetry-breaking in the system \cite{zoellner2020_influence,bloom2024chiral} from a vibrational point of view. 
	If CISS is driven by molecular vibrations, then it is expected that the spin selectivity $S(t)$ and the vibrational polarization $S_X(t)$ will be temporally correlated. 
	
	In order to quantify the relation between electronic and vibrational dynamics, 
	we introduce the total eigenstate population difference
	\begin{equation}
		\delta p = \frac{1}{2N}\sum_i^{2N} p_i [\Theta(\varepsilon_0 - \varepsilon_i) -\Theta(\varepsilon_i - \varepsilon_0)].\label{eq:delta_p}
	\end{equation}
	Here, $\Theta(x-y)$ is the Heaviside step-function and $\varepsilon_0$ the on-site energy of the electronic sites, which also marks the band center of the molecular electronic spectrum.
	This quantity measures the time-dependent difference of the occupancy of molecular electronic eigenstates with energies $\varepsilon_i<\varepsilon_0$ (below the center of the electronic spectrum) and the occupancy of states with energy $\varepsilon_i>\varepsilon_0$ (above the center). Here, the instantaneous eigenstate populations, $p_i(t)$, are the diagonal elements of reduced single-particle density matrix $\tilde{\rho}_\mathrm{el}(t)$ in the instantaneous eigenbasis of the electronic Hamiltonian including the electronic-vibrational interaction, $H_\mathrm{el}(X)$ (cf. \cref{eq:H_el_t}). That is, $\tilde{\rho}_\mathrm{el}(t) = U(t)\rho_\mathrm{el}(t)U^\dagger(t)$ is obtained from the instantaneous unitary transformation $U(t)$ that diagonalizes  $H_\mathrm{el}(X)$. By definition of the eigenstate populations $0\leq p_i\leq 1$, it is clear that $0 \leq \delta p\leq 1$. The derivation of the total population difference is given in Appendix \ref{app:A}.
	
	
	\section{Results}\label{sec:Results}
	In this section, we analyze the electronic and vibrational dynamics of a chiral molecular junction driven out of equilibrium using the Ehrenfest-HEOM approach. We start by discussing the parameters used in our calculations. Then, we explore the dynamics based on a single trajectory, focusing on how molecular vibrations affect the electronic spectrum, charge transport, and spin selectivity. This analysis helps us understand how bias voltage, system size, and electronic-vibrational coupling influence spin selectivity. Finally, we examine ensemble-averaged dynamics and consider the impact of temperature on spin selectivity.
	
	\subsection{Parameters}\label{sec:Results_A}
	
	Unless noted otherwise, we use the parameters introduced in Ref.~\cite{fransson2020vibrational}, which sets $\varepsilon_0 = -\SI{240}{\milli\electronvolt}$, $\omega_0=\SI{0.4}{\milli\electronvolt}$, $t_0=\SI{40}{\milli\electronvolt}$, $t_1 = \SI{4}{\milli\electronvolt}$, $\lambda_0 = \SI{1}{\milli\electronvolt}$, $\lambda_1 = \SI{0.1}{\milli\electronvolt}$, $\Gamma_L=\Gamma_R = \SI{10}{\milli\electronvolt}$, and $k_B T=\SI{25}{\milli\electronvolt}$. For simplicity, we also set the helix radius and pitch to be equal, $a=h=1$. 
	Since the thermal energy is much larger than the vibrational energy,  $k_B T\gg \hbar \omega$, and the dominant electronic energy scale is significantly larger than the vibrational energy, $t_0\gg \hbar \omega$, with a ratio of $\hbar\omega/t_0 = 0.01$, placing it in the adiabatic regime, a classical approximation of the vibrational modes is expected to be accurate. 
	
	\subsection{Single Trajectory Analysis}\label{sec:Results_B}
	We begin by considering a single trajectory with initial conditions $X_j(0) = 0$ and $P_j(0) = 0$ for all vibrational modes. These conditions reflect an uncharged molecule starting at equilibrium and decoupled from the leads. We first examine the system at a single voltage and then discuss how the results vary with different voltages.
	
	\subsubsection{Dynamics at a Fixed Voltage}\label{sec:Results_B_1}
	
\begin{figure}[!h]
		\centering
		\includegraphics[width=0.99\columnwidth]{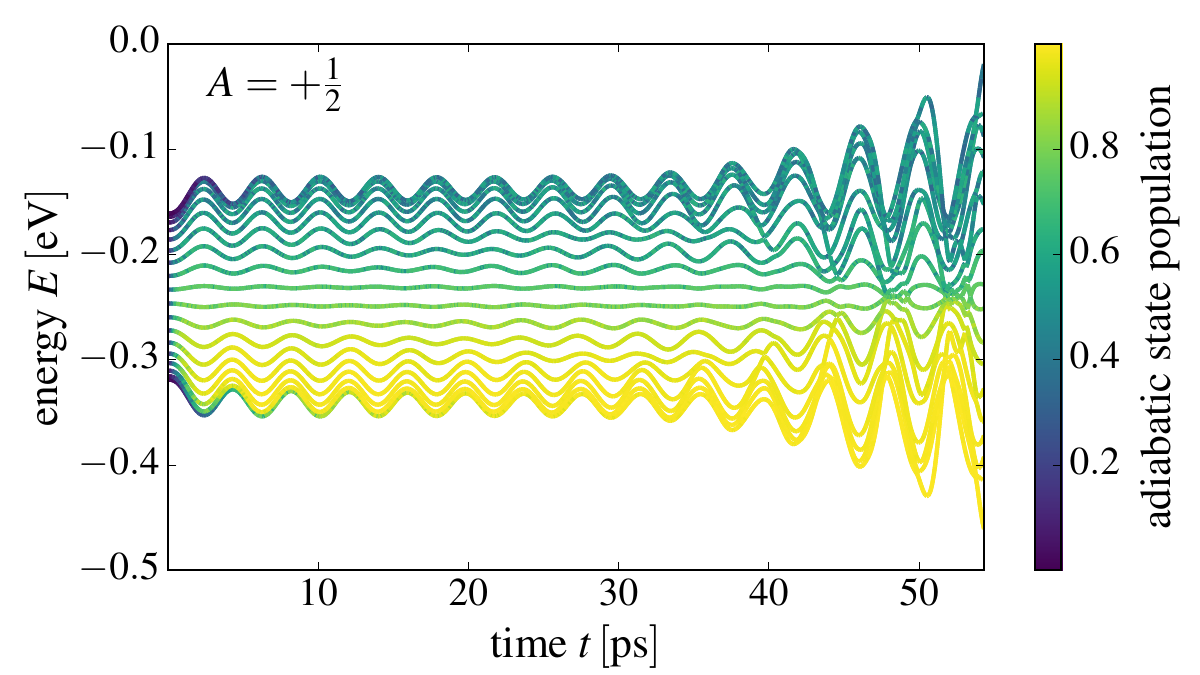}
		\caption{Time-dependent electronic spectrum (solid lines) and adiabatic state population (coloring of lines) for a system with size $N=3\times 6$ at bias voltage $\Phi=\SI{0.48}{\volt}$ and asymmetry parameter $A=+\frac{1}{2}$. }\label{fig:Results_1_1}
	\end{figure}

	We focus on a bias voltage of $\Phi=\SI{0.48}{\volt}$, where the chemical potential of the right lead aligns with the molecular band center at $E=\varepsilon_0 =-\SI{0.24}{\electronvolt}$. \Cref{fig:Results_1_1} illustrates the time-dependent electronic spectrum (solid lines), obtained by diagonalizing the electronic Hamiltonian $H_\mathrm{el}(t)$ at each time step for a system with size $N=3\times 6$ and asymmetry parameter  $A=+\frac{1}{2}$. The coloring indicates the occupation of these electronic eigenstates, represented by the corresponding populations, $p_i$. 
	
	Upon applying a voltage of $\Phi=\SI{0.48}{\volt}$, all electronic states of the molecule become at least partially occupied. States with energy below $E\leq \SI{0.24}{\electronvolt}$ are nearly fully populated, as they are energetically below the right lead chemical potential. Additionally, the electronic spectrum exhibits an oscillatory behavior over time, with the oscillation amplitude increasing until it reaches saturation, although this occurs beyond the time interval shown here. During this oscillation, low-energy states temporarily come closer to the center of the electronic spectrum, such that even these contribute to some degree to electronic transport as their population changes over time. This temporal behavior can be attributed to the dynamics of molecular vibrations; specifically, the increasing amplitude of the molecular vibrational coordinates is reflected by the increased amplitude of oscillations of the electronic spectrum. 
	
	It is important to note that the instantaneous eigenstates are two-fold degenerate due to Kramers theorem since the molecular Hamiltonian within the fixed-nuclei approximation is time-reversal symmetric. Consequently, when neglecting feedback from the vibrational dynamics, no CISS effect is observed in the purely electronic transport calculations \cite{kramers1930theorie,wigner1993operation}. 
	\begin{figure}[!h]
		\centering
		\includegraphics[width=0.99\columnwidth]{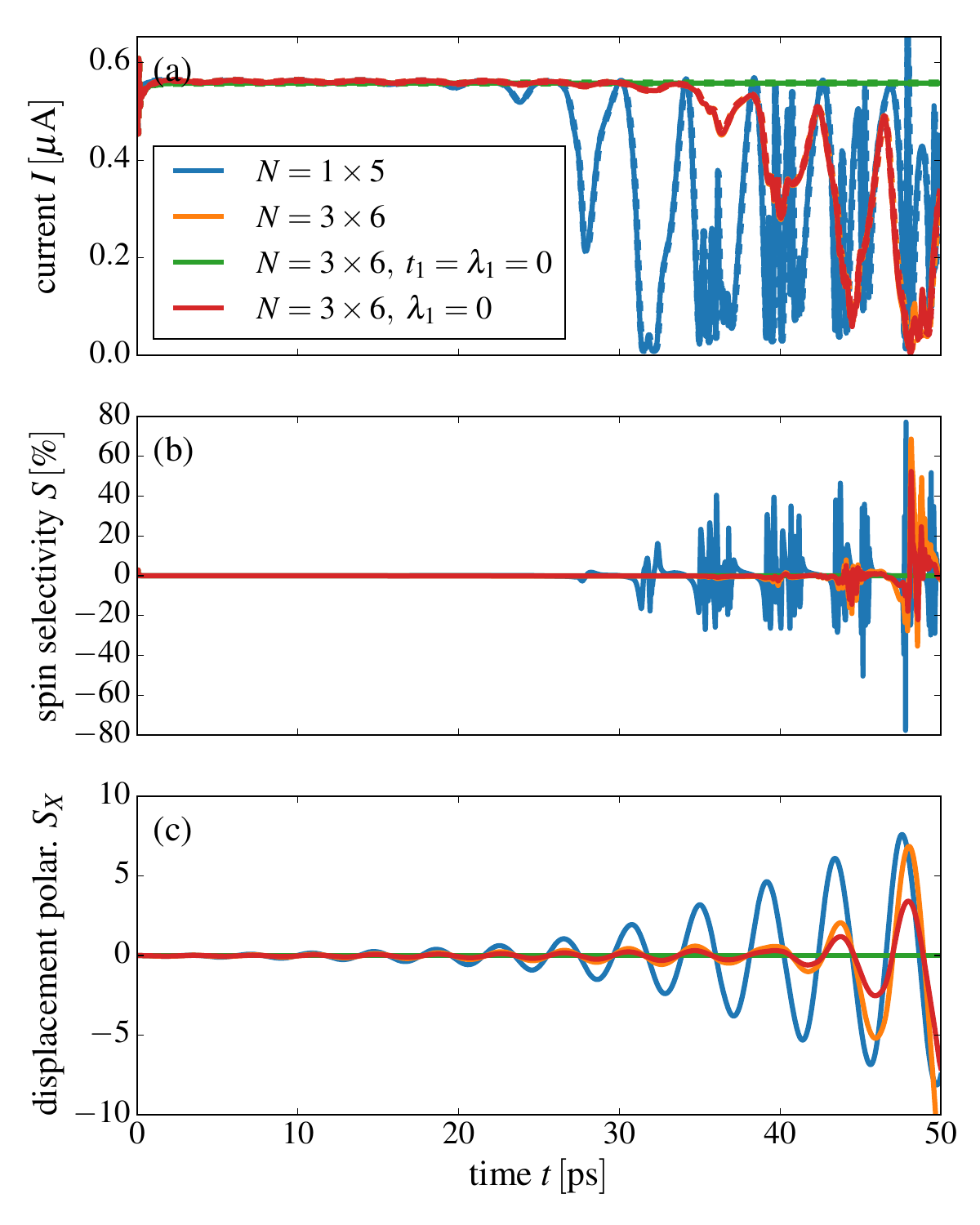}
		\caption{(a) Dynamics of charge current $I$, (b) spin selectivity $S$, and (c) displacement polarization $S_X$, for system sizes $N=1\times 5$ (black lines), $N=3\times 6$ (red lines) and $N=3\times 6$ without electronic-vibrational coupling (blue lines). 
			In (a), solid lines show currents for $A=+\frac{1}{2}$ and dashed lines for $A=-\frac{1}{2}$. }\label{fig:Results_1}
	\end{figure}
	
	Next, we examine the spin selectivity and displacement polarization for various system sizes. \Cref{fig:Results_1} showcases the dynamics of various observables and the parameters stated in Section \ref{sec:Results_A} for system sizes $N=1\times 5$ (blue) and $N=3\times 6$ (orange), and for a fixed system size of $N=3\times 6$ considering no  electronic-vibrational couplings $t_1=\lambda_1=0$ (green) and setting only the vibrationally-assisted SOC to zero $\lambda_1=0$ (red). 
	
	Initially, the short-time charge current dynamics, shown in \cref{fig:Results_1} (a), is very similar for both system sizes and opposite lead magnetizations, where $A=+\frac{1}{2}$ is indicated by solid lines and $A=-\frac{1}{2}$ by dashed lines. However, at longer times, a distinct modulation of the current is observed. The amplitude of this modulation grows with time, eventually leading to a notable suppression of the charge current at longer times, particularly for $t>\SI{25}{\pico\second}$. As with the oscillations in the electronic spectrum, these oscillations saturate at longer times. This phenomenon is attributed to large vibrational displacements that significantly modify the electronic spectrum of the molecule, as the displacement of the inter-site vibrational modes modulates the effective hopping amplitude $-t_0+t_1X_j$ between neighboring sites $j$ and $j+1$ in the molecule. 
	Since, for a finite voltage bias, electron transport through the molecule is predominantly unidirectional, suppression of electronic hopping leads to a suppression of charge current. 
	
	Particularly at times when the charge current is strongly modulated by molecular vibrations, a substantial but transient spin selectivity of up to $\pm 80\%$ is observed, as shown in \cref{fig:Results_1} (b). The onset of the charge current suppression and large but transient spin selectivity occurs at later times for the larger system, indicating that the molecular length plays a critical role in the stability and dynamical behavior of the system under an applied voltage bias. The transient spin selectivity is evidently correlated with the displacement polarization, as one sees in \Cref{fig:Results_1} (c) where the peaks in $S(t)$ and $S_{X}$ occur at the same times.  
	
	Turning to the influence of electronic-vibrational couplings, we find that for the system without electronic-vibrational coupling ($t_1=\lambda_1=0$), the charge current reaches a stationary state after $t\sim \SI{1}{\pico\second}$. 
	As expected, the oscillations of the current present in the system with electronic-vibrational coupling are completely absent in the system without such couplings. In this case, the spin selectivity is negligibly small, $S < 10^{-6}\%$, and the displacement polarization vanishes due to the absence of electron-driven vibrational dynamics. This is in stark contrast to the case with $t_1=\SI{4}{\milli\electronvolt}$ and $\lambda_1=0$ (red lines in \cref{fig:Results_1}), where the spin-dependent electronic-vibrational coupling is still zero and for which the charge currents are very similar to the ones for the original parameters (orange lines). 
	
	These results show that a finite spin selectivity cannot be solely attributed to the large displacement of vibrational modes alone.  
	To achieve a finite spin selectivity, the molecular vibrational modes need to break the symmetry of the charge current with respect to switching of the left leads magnetization, meaning bond displacements must differ sufficiently for opposing lead magnetizations. This is evidenced by the correlation between the dynamics of $S_X$ and $S$. 
	
	\subsubsection{Voltage-Dependence of Electronic and Vibrational Dynamics}\label{sec:Results_B_2}
	
\begin{figure}[!h]
		\centering
		\includegraphics[width=0.99\columnwidth]{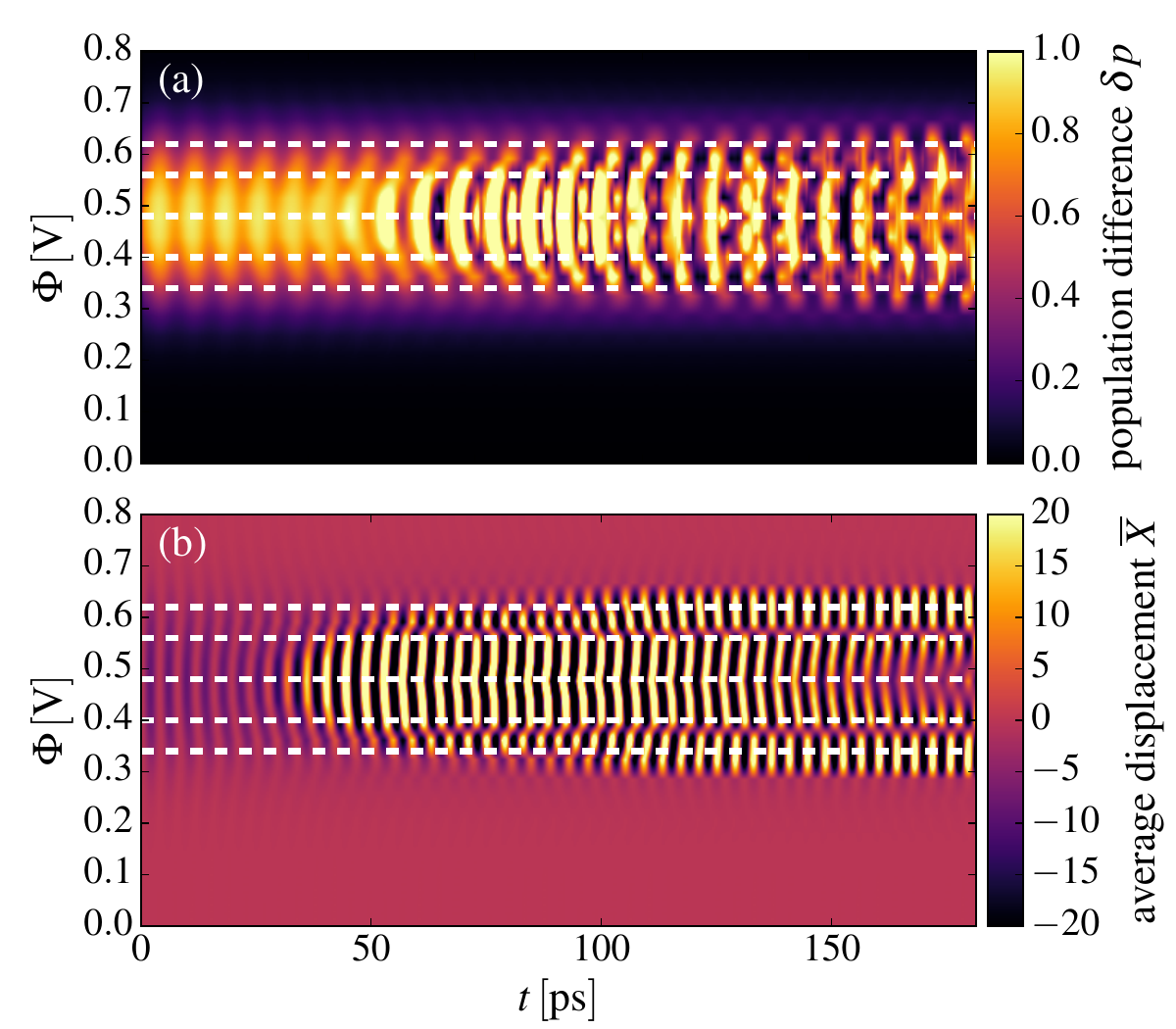}
		\caption{(a) Dynamics and voltage dependence of population difference $\delta p$, and (b) dynamics and voltage dependence of average vibrational coordinate $\bar{X}$. The asymmetry is $A=\frac{1}{2}$, and the system size is $N=1\times 5$. White dashed lines denote the bare eigenenergies of the electronic Hamiltonian.}\label{fig:Results_2_a}
	\end{figure}

	Next, we investigate the voltage dependence of the dynamics for a system of size $N=1\times 5$ and lead polarization $A=+\frac{1}{2}$. In \cref{fig:Results_2_a} (a), we show the voltage and time dependence of the population difference, $\delta p$ (cf. \cref{eq:delta_p}). 
	To quantify the corresponding vibrational dynamics, \cref{fig:Results_2_a} (b) also depicts the voltage-dependence of the average vibrational coordinate, $\bar{X}$.
	
	We first discuss the population difference. In the voltage range $\SI{0.34}{\volt} \leq \Phi \leq \SI{0.62}{\volt}$, the population difference increases rapidly, as electronic states are populated. 
	In this voltage regime, only part of the bare molecular electronic eigenspectrum, denoted by white dashed lines, is in the bias window, such that the population difference is generally nonzero. 
	In contrast, for voltages $\SI{0}{\volt}\leq \Phi< \SI{0.34}{\volt}$, all eigenstates are below the bias window and all states are fully populated, $p_i = 1$; consequently, the population difference vanishes. Similarly, for voltages $\Phi\geq \SI{0.62}{\volt}$, all electronic states are in the bias window and the electronic population converges to the infinite-bias limit of the resonant level model with a population of $p_i=\frac{1}{2}$ of all states \cite{haug2008quantum}. 
	This means $p_i = 0$ and $\delta p = 0$. 
	
	With this in mind, we focus on the voltage regime for which $\delta p$ is nonzero. After the initial increase, the population difference exhibits clear oscillations over time. The amplitude and frequency of these oscillations are voltage-dependent. This oscillatory behavior reflects the interplay between electronic and vibrational dynamics, as is seen in \Cref{fig:Results_2_a} (b).
	Here, the voltage-dependent dynamics of the average displacement $\bar{X}$ also shows sustained time-dependent oscillations, similar to $\delta p$. The frequency and amplitude of oscillation is influenced by the applied voltage, highlighting the dynamic response of the molecular structure to electronic excitations. With increasing time, the oscillation amplitude of $\bar{X}$ grows, indicating an increasing energy transfer from the electronic to vibrational degrees of freedom, eventually reaching a saturation point. Outside of this voltage regime, the average displacement vanishes and $\bar{X}=0$. This can be explained by vanishing electronic forces, indicated by $\delta p = 0$. In this voltage regime, vibrational dynamics is purely determined by the harmonic potential and the initial condition, which, for these calculations, is $X_j(0)=P_j(0)=0$.
	
	The similarity in temporal patterns of $\delta p$ and $\bar{X}$ emphasizes the strong coupling between electronic dynamics and molecular vibrations.  
	From this analysis, we can deduce that when the center of the molecular electronic spectrum is equal to the chemical potentials of the leads in equilibrium, which is achieved by setting the on-site energies to $\varepsilon_0 = 0$, there is no prominent vibrational dynamics for any voltage. Hence, in this case, spin selectivity vanishes for all voltages within the Ehrenfest approach. 
	\begin{figure}[!h]
		\centering
		\includegraphics[width=0.99\columnwidth]{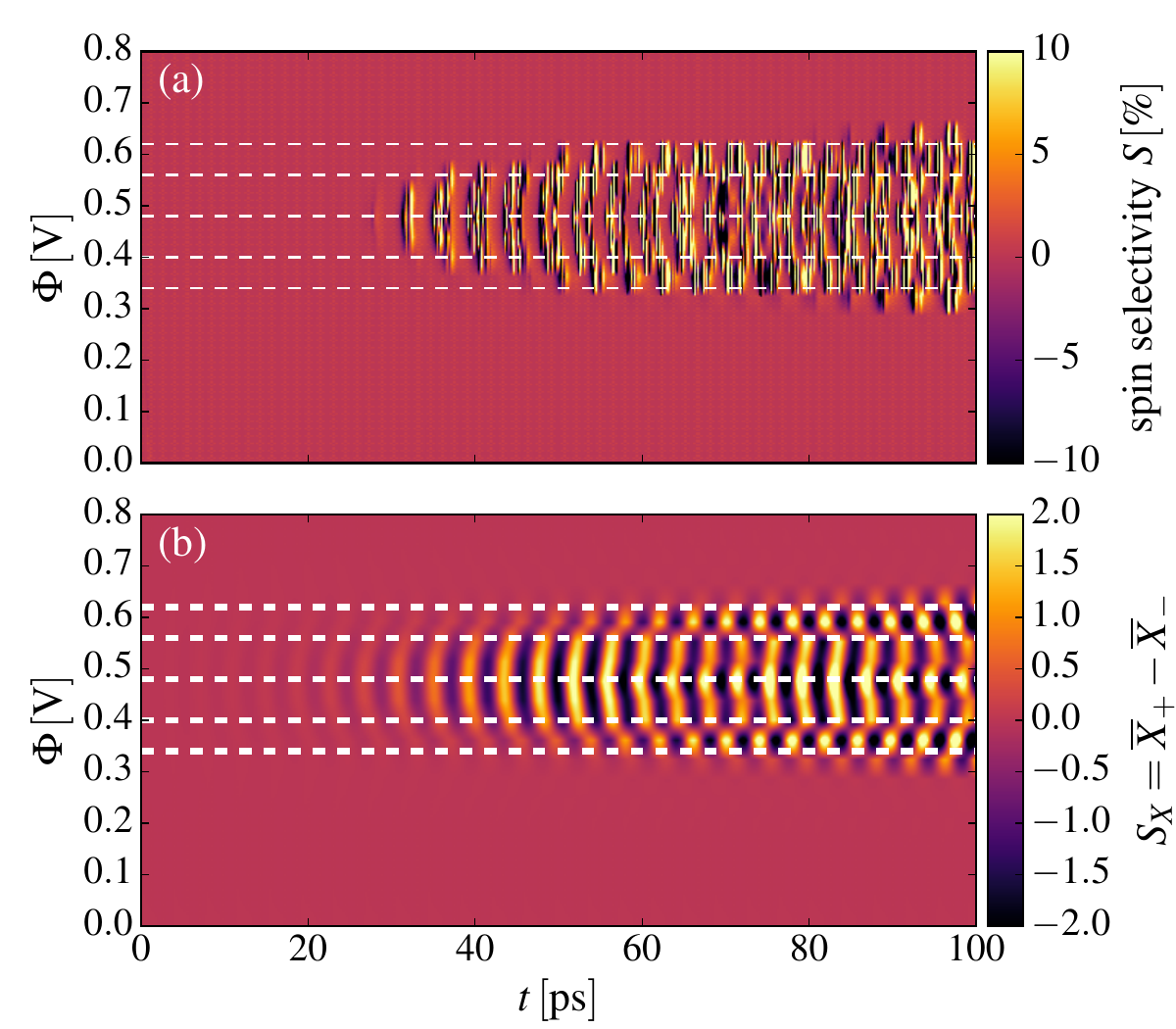}
		\caption{(a) Dynamics and voltage dependence of spin selectivity $S$ and (b) vibrational polarization $S_X=\bar{X}_+-\bar{X}_-$ for a system with size $N=1\times 5$. Similar to \cref{fig:Results_2_a}, white dashed lines denote the bare eigenenergies of the electronic Hamiltonian.}\label{fig:Results_2_b}
	\end{figure}
	
	Next, we examine the combined voltage and time dependence of the spin selectivity displacement polarization, which are shown in \cref{fig:Results_2_b} (a) and (b). Similar to the voltage-dependence of the population difference and the average displacement shown in \cref{fig:Results_2_a}, the spin selectivity and the displacement polarization are finite only in the voltage regime $\SI{0.34}{\volt} \leq \Phi\leq \SI{0.62}{\volt}$ which can be explained with the same reasoning as before.
	Both quantities have a strong voltage-dependent oscillation frequency and amplitude. 
	Similar to the comparison presented in Section~\ref{sec:Results_B_1}, peaks in the spin selectivity are temporarily correlated with maxima in the displacement polarization across the entire voltage range of $\SI{0.34}{\volt} \leq \Phi \leq \SI{0.62}{\volt}$.
	The influence of other parameters is discussed in detail in Appendix~\ref{app:B}, indicating, that the magnitude of spin selectivity is mainly determined by the static SOC $\lambda_0$ and the spin-independent electronic-vibrational coupling $t_1$.
	
	Finally, we note that the results discussed in this section show that the dynamics of a single trajectory are marked by strong oscillations at long times.  
	It is well known that Ehrenfest dynamics predict self-sustained Van der Pol oscillations of the vibrational modes. 
	A common explanation is that these oscillations arise as the coordinates evolve into a region characterized by negative friction \cite{hopjan2018molecular, bode2011scattering, bode2012current, kartsev2014nonadiabatic}. 
	
	\subsection{Analysis of Ensemble-Averaged Observables}\label{sec:Results_C}
	
	The Ehrenfest approach reveals that a single trajectory in this system does not reach a stationary state for the parameters investigated. The analysis is still useful in the sense that it helps identify the couplings responsible for spin selectivity. To obtain physically meaningful results from Ehrenfest calculations, however, it is necessary to sample the initial vibrational phase-space coordinates condition from the classical limit of the corresponding quantum initial condition and calculate an ensemble-averaged observables. In this section, we discuss the results of such an approach. 
	We demonstrate that employing this averaging procedure, the long-time oscillations observed for individual trajectories disappear, and the system reaches a stationary state. 
	
	\subsubsection{Initial Condition and Ensemble Averaging} 
	
	In the following calculations, we assume that all electronic states of the molecule are initially unoccupied. To obtain an approximation to the full quantum dynamics, the wave packet motion of the vibrational degrees of freedom is mimicked by propagating a swarm of individual trajectories starting with initial vibrational coordinates and momenta sampled from a distribution. This is the standard quantum-classical Ehrenfest approach to molecular dynamics \cite{mclachlan1964variational,micha1983self,tully1998mixed,kirrander2020ehrenfest}. Under the assumption that the electronic configuration is initially unoccupied, the vibrational modes are in the thermal Gibbs state determined by the harmonic potential in $H_\mathrm{vib}$. By Wigner transforming the thermal Gibbs state, one obtains a probability distribution for the initial coordinates and momenta, given by
	\begin{equation}
		\rho_\mathrm{W}^\mathrm{G}(X,P) = \prod_j 2\alpha_j \exP{-\alpha_j (X_j^2+P_j^2)},\label{eq:Wigner_dist}
	\end{equation}
	where $\alpha_j = \tanh\left(\frac{\beta \hbar \omega_j}{2}\right)$. In the following, ensemble-averaged observables $O$ (both quantum and classical) are denoted by $\langle O\rangle$, where
	\begin{equation}
		\langle O(t)\rangle =\int \dif X \dif P \rho_\mathrm{W}^\mathrm{G}(X,P) O(t).\label{eq:Ensemble_averaged_obs}
	\end{equation}
	Here, $O(t)$ are the propagated single trajectory observables and $\dif X=\prod_j \dif X_j, \dif P=\prod_j \dif P_j$. In practice, the initial coordinates and momenta $X_j(0), P_j(0)$ are drawn from the distribution \cref{eq:Wigner_dist} and then propagated along with the electronic density matrix and the auxiliary density matrices. Results are then obtained as an average over many single-trajectory calculations.
	
	\subsubsection{Dynamics at a Fixed Voltage}\label{sec:RC_2}
	
\begin{figure}[!h]
		\centering
		\includegraphics[width=0.99\columnwidth]{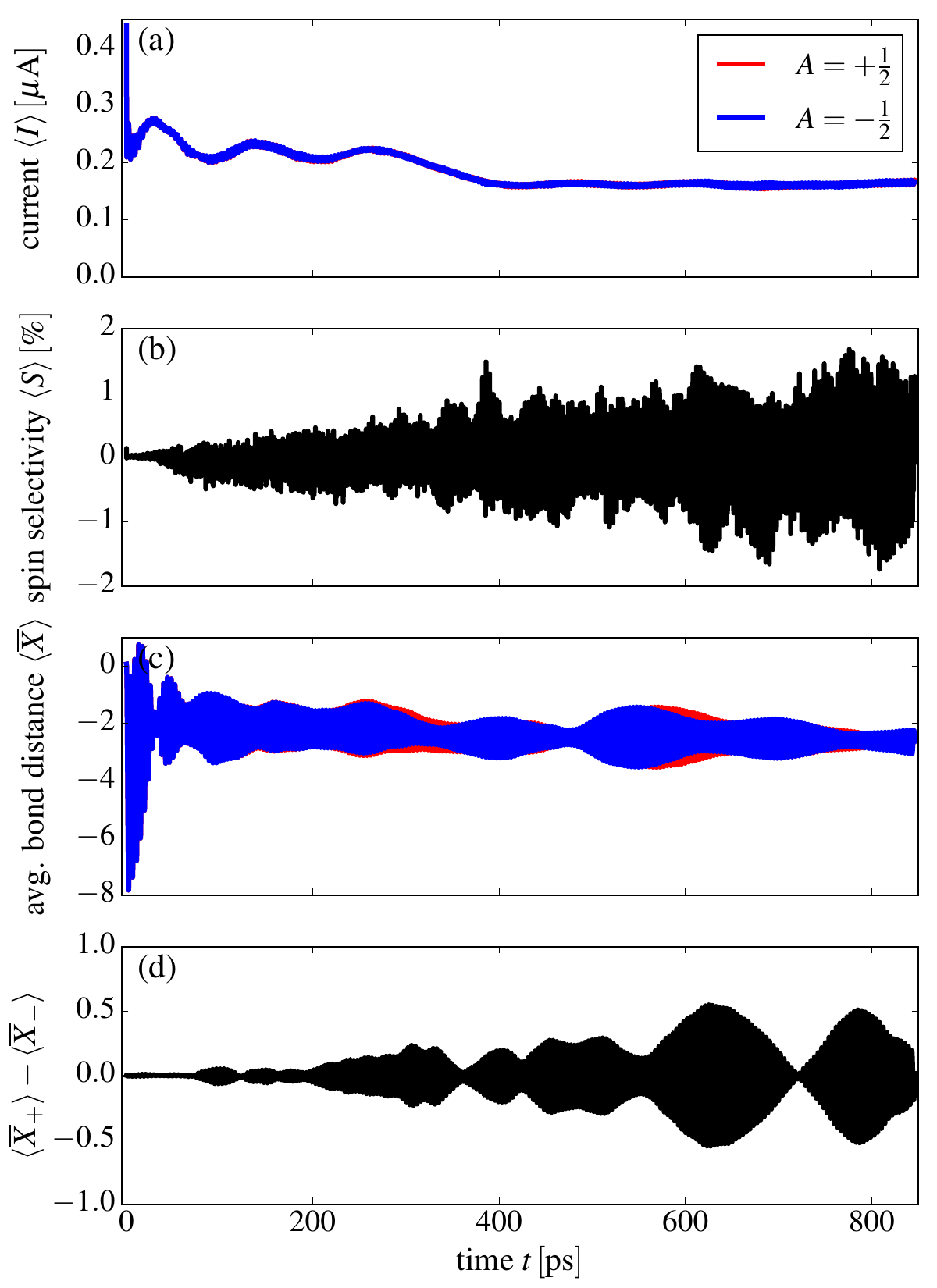}
		\caption{(a) Dynamics of ensemble averaged charge current $\langle I\rangle$ for lead polarizations $A=\frac{1}{2}$ (red) and $A=-\frac{1}{2}$ (blue), (b) spin selectivity, (c) average bond distance $\langle \bar{X}\rangle$, and (d) displacement polarization $\langle \bar{X}_+\rangle - \langle \bar{X}_-\rangle $ (d). Results are obtained from an average over $12000$ trajectories. System size is $N=1\times 5$ sites and the voltage is $\Phi=\SI{0.48}{\volt}$.}\label{fig:Results_C_1}
	\end{figure}

	\Cref{fig:Results_C_1} shows the dynamics of ensemble averaged observables for a system with $N=1\times 5$ sites at a voltage bias $\Phi=\SI{0.48}{\volt}$. 
	This voltage bias has been chosen because it gives a particularly large single-trajectory spin selectivity, which can be seen in \cref{fig:Results_2_b} (a)). Other parameters are as in the previous section and the ensemble average is taken over 12000 trajectories.
	Convergence of results has been confirmed and is discussed in detail in Appendix \ref{app:C}.
	\Cref{fig:Results_C_1} (a) shows the dynamics of the charge currents $\langle I_\pm\rangle$ for $A=+\frac{1}{2}$ (red), and $A=-\frac{1}{2}$ (blue) and \cref{fig:Results_C_1} (b) the spin selectivity $\langle S\rangle$, obtained from the ensemble-averaged charge currents $\langle I_+\rangle$ and $\langle I_-\rangle$. Because the spin selectivity is not an independent observable, we do not consider the average spin selectivity in the sense of a mean of single-trajectory spin selectivities, but rather the spin selectivity obtained from the mean currents,
	\begin{equation}
		\langle S\rangle(t) = \frac{\langle I_+(t)\rangle - \langle I_-(t)\rangle}{\langle I_+(t)\rangle+\langle I_-(t)\rangle}.
	\end{equation}
	To quantify the vibrational dynamics, in \cref{fig:Results_C_1} (c) the dynamics of the ensemble-averaged average bond distance $\langle \bar{X}_p\rangle$, for $A=\frac{1}{2}$ (red) and $A=-\frac{1}{2}$ (blue), and in \cref{fig:Results_C_1} (d) the ensemble-averaged displacement polarization $\langle \bar{X}_+\rangle - \langle \bar{X}_-\rangle$ is depicted.
	
	The ensemble-averaged electronic observables reach a nonequilibrium stationary state at $\sim \SI{400}{\pico\second}$ (\cref{fig:Results_C_1} (a)), contrary to the dynamics of single trajectories. Similarly, the vibrational observables also reach a stationary state, with \cref{fig:Results_C_1} (c) showing a long-time limit of $\langle \bar{X}_\pm \rangle < 0$, indicating that the molecule is on average contracted with respect its equilibrium configuration. 
	This can be explained with a similar analysis as that given in Sec.~\ref{app:A} for a simplified two-site Peierls model. 
	Here, one finds that occupying all states below the molecular band center, such that $p_-=1$ and $p_+=0$, leads to a shifted potential of mean force, with its minimum at a contracted molecular geometry.
	
	Although there are still small oscillations of the charge current at longer times, $t\geq \SI{400}{\pico\second}$, we attribute these to the finite number of trajectories in the calculations. Tests indicate that these oscillations decrease with increasing trajectory number (data not shown). These oscillations lead in turn to small deviations in the currents $\langle I_+\rangle$ and $\langle I_-\rangle$, which results in an oscillation of the spin selectivity around $S=0$ at later times. The spread of the spin selectivity, however, is mainly determined by the finite number of trajectories. Therefore, we provide an upper and lower bound on the spin selectivity in the trajectory average of around $S\sim \pm 2\%$. This is significantly smaller than the spin selectivity observed for a single trajectory. Similarly, the difference between the ensemble-averaged mean vibrational displacements are significantly smaller, leading to a strong reduction of the vibrational polarization (\cref{fig:Results_C_1} (c) and (d)). 
	
\begin{figure}[!h]
		\centering
		\includegraphics[width=0.99\columnwidth]{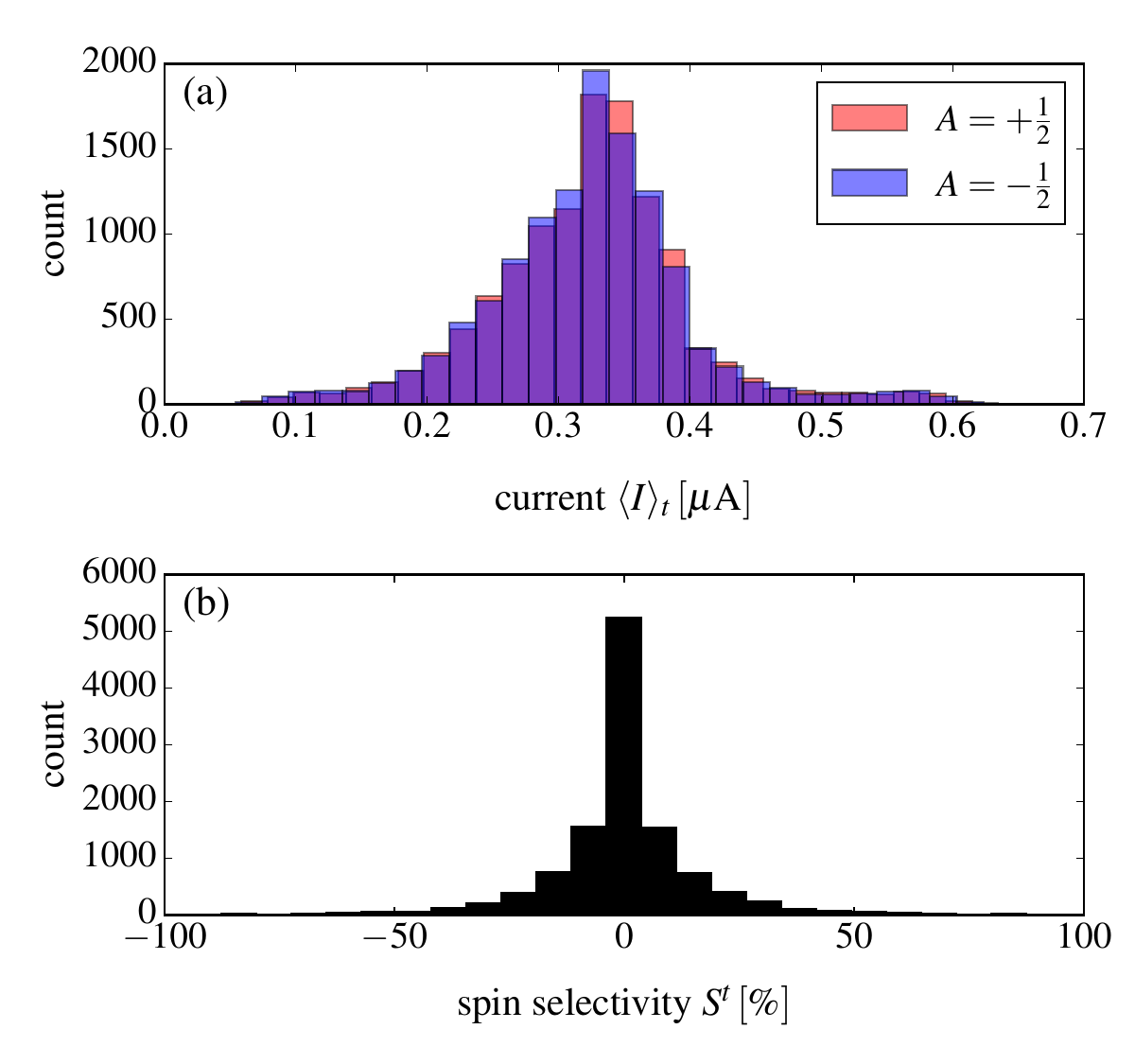}
		\caption{Ensemble distribution of time-averaged charge currents (a) (red for $A=+\frac{1}{2}$, blue for $A=-\frac{1}{2}$), spin selectivity obtained from single trajectories (b). In total, 12000 trajectories have been used in this analysis. Parameters are described in the main text.}\label{fig:Results_C_2}
	\end{figure}

	For further analysis, we investigate the ensemble statistics of charge currents and spin selectivities, shown in \cref{fig:Results_C_2} in panels (a) and (b), respectively. For this, we compute the time-average of single-trajectory charge currents within the time-interval $t\in [400,800]\SI{}{\pico\second}$, denoted by $\langle I\rangle_t$. This is the time interval at which the system is in a stationary state (see \cref{fig:Results_C_1}). 
	For each trajectory, we compute the spin selectivity from the time-averaged charge currents,
	\begin{equation}
		S^t = \frac{\langle I_+\rangle_t - \langle I_-\rangle_t}{\langle I_+\rangle_t+\langle I_-\rangle_t}.
	\end{equation} 
	
	From the ensemble statistics of time-averaged charge currents (\Cref{fig:Results_C_2} (a)), it appears that time-averaged currents for opposite lead magnetizations ($A=+\frac{1}{2}$ (red) and $A=-\frac{1}{2}$(blue)) are very similar, although the charge current dynamics for single trajectories was quite different for opposite lead magnetizations. This is also reflected in the broad ensemble distribution of the spin selectivity of approximately $\pm 50\%$, with about an equal number of trajectories contributing to a positive and negative spin selectivity. 
	This broad range of time-averaged charge currents and spin selectivities is a result of the interplay between electron and vibrational dynamics, as in the absence of electronic-vibrational coupling, the ensemble distribution of charge currents would have a vanishing standard deviation.
	We expect that similar results are obtained for other voltages in the regime, in which a few states are in the bias window, as discussed in Section~\ref{sec:Results_B}.
	
	\subsubsection{Temperature Dependence of Spin Selectivity}\label{sec:Results_C2}
	
	The finite width of the ensemble distributions discussed in the previous section suggests an influence of the initial thermal probability distribution on the final distribution of ensemble-averaged observables. In order to study this influence, we analyze the ensemble distributions of time-averaged spin selectivities for $k_B T=\SI{7}{\milli\electronvolt}$ (corresponding to $T=\SI{81}{\kelvin}$) and compare it to the distribution for $k_BT =\SI{25}{\milli\electronvolt}$ as in the previous analysis. Here, we consider a system with size $N=3\times 7$, and in the ensemble statistics a total of $1000$ trajectories have been used. We expect the results to be representative, also for other parameters.
	\begin{figure}[!h]
		\centering
		\includegraphics[width=0.99\columnwidth]{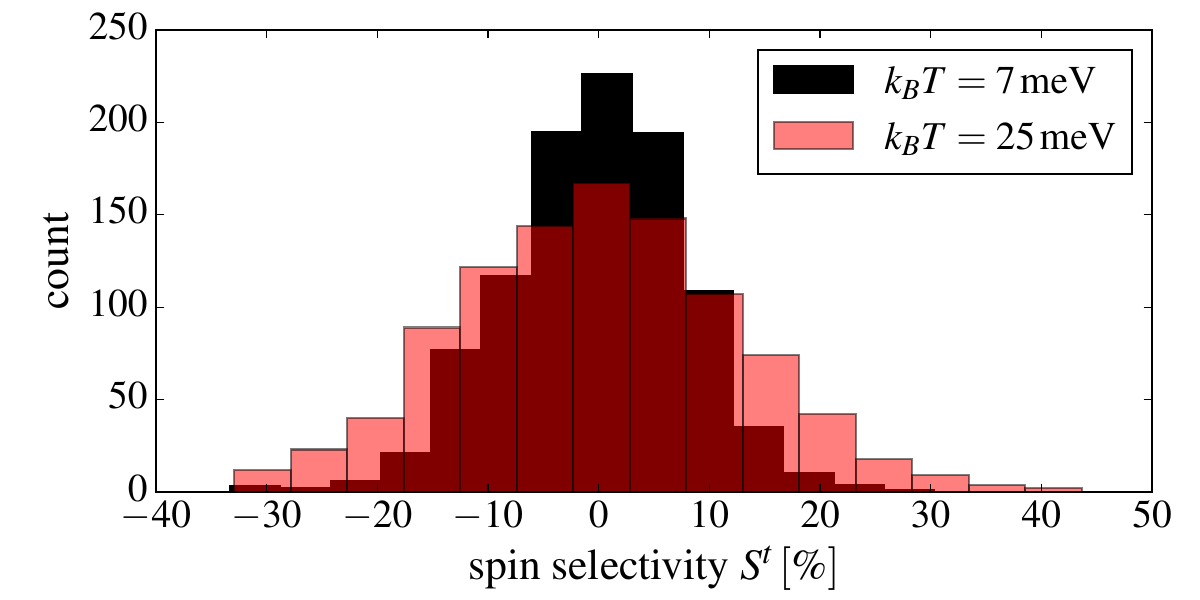}
		\caption{Comparison of ensemble distribution of spin selectivity for temperatures $k_B T=\SI{7}{\milli\electronvolt}$ (black) and $k_B T=\SI{25}{\milli\electronvolt}$ (red) at a system of size $N=3\times 7$ at voltage bias $\Phi=\SI{0.48}{\volt}$. The statistics have been collected from 1000 trajectories.}\label{fig:Res_C_3}
	\end{figure}
	
	The results depicted in \cref{fig:Res_C_3} show that reducing the temperature leads to a clear reduction of the width of the ensemble distributions. 
	Consequently, numerous trajectories exhibit a finite spin selectivity in the time average.  Surprisingly, spin selectivity nearly vanishes completely in the ensemble average, which is in stark contrast to the single-trajectory results and to the ones reported in Ref.~\cite{fransson2020vibrational}, in which spin selectivities of approximately $S\sim 12\%$ are reported for the corresponding system size and bias voltage.
	However, some of the trajectories contribute to positive spin selectivity, whereas other trajectories to negative spin selectivity. 
	In the ensemble average, these individual contributions with finite spin selectivity balance each other out leading to a nearly vanishing spin selectivity. 
	This explains the discrepancy between single trajectory and ensemble average results.

	\section{Conclusion}\label{sec:Conclusion}
	
	We have analyzed the effect of molecular vibrations on the chirality-induced spin selectivity effect using a mixed quantum-classical approach. 
	The results show that finite spin selectivities correlate with vibrational dynamics in single trajectories. To quantify this correlation, we have introduced a figure of merit, termed displacement polarization, which measures the difference in average molecular displacements for opposite lead magnetizations.
	This figure of merit mirrors the voltage-dependent, nonequilibrium dynamics of spin selectivity, strongly indicating that vibrational dynamics are a key factor in driving CISS.
	
	However, while significant spin selectivity is observed in single trajectories, the ensemble average reveals an almost vanishing selectivity. This is due to the symmetric distribution of spin selectivities around zero when averaging over multiple trajectories. Thus, the effect observed in individual trajectories is masked in ensemble and time-averaged results.
	
	Our findings indicate that while vibrational dynamics can induce spin selectivity, the effect is diminished in ensemble averages within a mean-field treatment. This suggests the need for further studies beyond mean-field treatments to fully understand and confirm the role of nonequilibrium dynamics of molecular vibrations in CISS, which is the focus of ongoing research.
	
	\section*{Acknowledgements}\label{sec:Acknowledgement}
	
	R.S. and M.T. acknowledge support by DFG-funded Research Training Group DynCAM (Grant No. RTG 2717). S.L.R. thanks the Alexander von Humboldt Foundation
	for the award of a Research Fellowship. Furthermore, the authors acknowledge the support by the state of Baden-Württemberg through bwHPC and the DFG through Grant No. INST 40/575-1 FUGG (JUSTUS 2 cluster).
	\appendix
	\section{Derivation of the Population Difference}\label{app:A}
	
	The motivation for the functional form of the eigenstate population difference $\delta p$, which is given in \cref{eq:delta_p}, is rooted in the following. 
	When the spin-independent electronic-vibrational coupling is the dominant interaction, $t_1 \gg \lambda_1$, one can consider only this contribution to the total electronic-vibrational interaction Hamiltonian $H_\mathrm{el-vib}$. 
	The system with two sites only with a single inter-site vibrational mode is described by the Hamiltonian
	\begin{equation}
		H(X) = -\frac{t_1}{\sqrt{2}} X (c_1^\dagger c_2 + \mathrm{h.c.}).
	\end{equation}
	The spectrum of this model in the single-particle basis is given by
	\begin{equation}
		\varepsilon_\pm = \pm \left\vert \frac{t_1}{\sqrt{2}}X\right\vert, \qquad \psi_\pm = \frac{1}{\sqrt{2}}\begin{pmatrix}
			\mp \mathrm{sgn}(X)\\ 1
		\end{pmatrix}.  
	\end{equation}
	Here, $\mathrm{sgn}(X)$ is the sign function of the vibrational coordinate $X$. For this toy model within the quantum-classical approximation, the potential of mean force takes the form
	\begin{equation}
		U_\mathrm{MF}(X) = \frac{\omega_0 X^2}{2} + \frac{t_1}{\sqrt{2}} X (p_- - p_+),
	\end{equation}
	which demonstrates that the electronic-vibrational interaction contributes to the mean force via the difference in populations of states below and above the molecular band center: $(p_- - p_+)$. Note, that, by virtue of the fixed-nuclei approximation used in this analysis, $p_+$ and $p_-$ are the adiabatic populations when neglecting coupling to the leads. 
	
	In order to generalize this explanation to the chiral model used in this work, the populations $p_i$ have to be evaluated in the instantaneous eigenbasis of the electronic Hamiltonian of the molecule, $H_\mathrm{el}(X) = H_\mathrm{el}^{(0)}+H_\mathrm{el-vib}(X)$ (cf. \cref{eq:H_el_t}), for all times. 
	Since the electronic-vibrational interactions of the chiral model are all within the molecular part of the total Hamiltonian, the electronic force calculated for the true model will still only depend on population differences within the molecule and not on states in the lead, although these of course still influence molecular electronic eigenstates. 
	
	With these considerations in mind, $\delta p$ can be understood as a generalization of the electronic contribution due to the Peierls-coupling to the force of electrons in a multi-level system on a collective vibrational displacement $\bar{X}$ by making the ansatz 
	\begin{align}
		H &=  \sum_j -\frac{t_1}{\sqrt{2}}X_j (c_j^\dagger c_{j+1}^\pdagger + \mathrm{h.c.})\\
		&\sim \bar{X} \sum_k \varepsilon_k^\pdagger c_k^\dagger c_k^\pdagger, 
	\end{align} 
	where $\varepsilon_k$ are the single particle eigenenergies of the bare electronic system (that is for the case $X_j=0$). Although this separation is not exact, it still provides a simple measure for the force acting on the vibrations, generated by the electrons. 
	In a sense, this separation neglects electronic coherences over more than two sites.
	The motivation behind the proposed functional form of the eigenstate population difference $\delta p$ defined in \cref{eq:delta_p} is based on considering only the spin-independent electronic-vibrational interaction of the reduced two-site model. 
	\begin{figure}[!h]
		\centering
		\includegraphics[width=0.98\columnwidth]{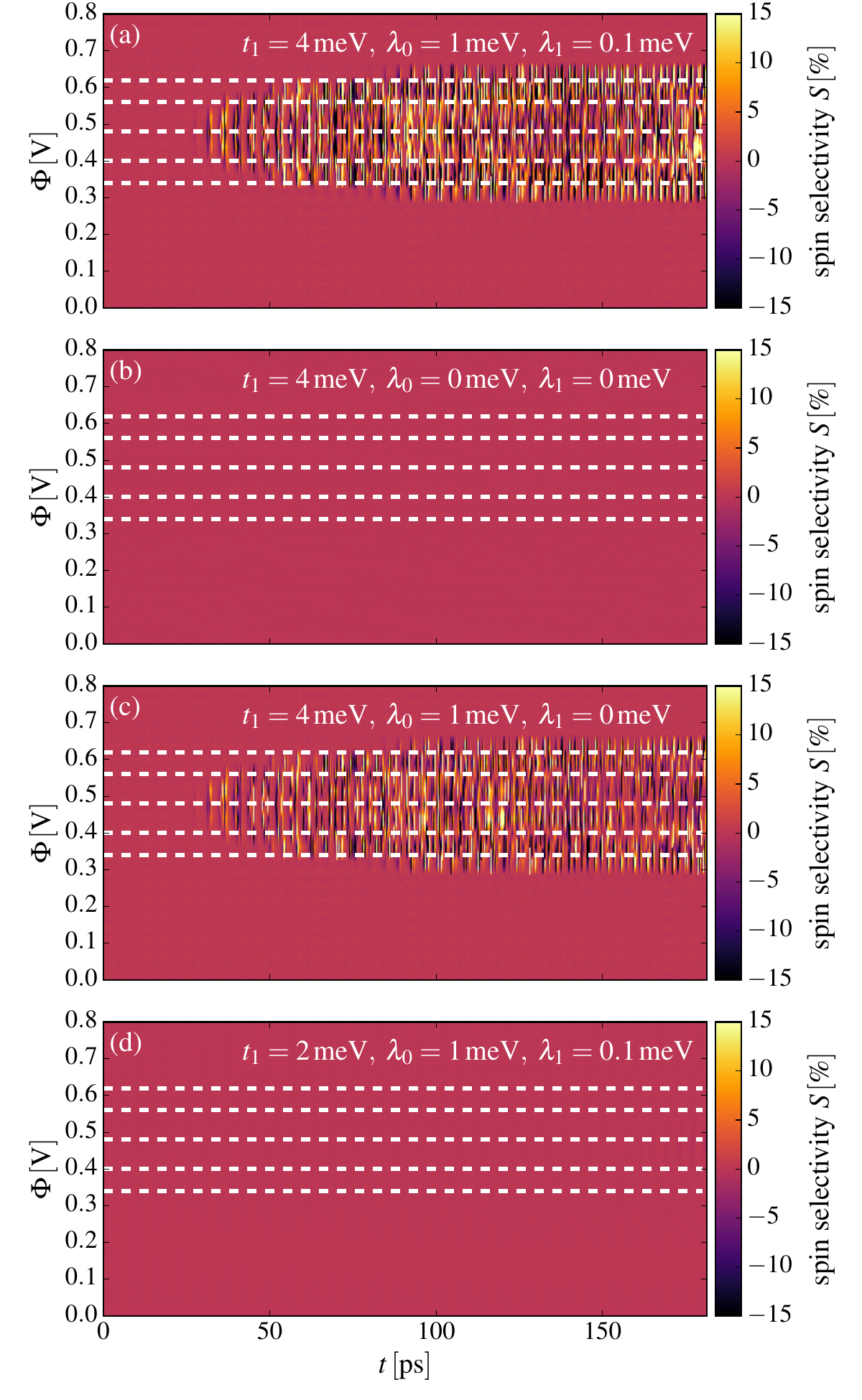}
		\caption{Dynamics and voltage-dependence of spin selectivity for a system size $N=1\times 5$. (a) $t_1=\SI{4}{\milli\electronvolt},\lambda_0=\SI{1}{\milli\electronvolt},\lambda_1=\SI{0.1}{\milli\electronvolt}$, (b) $t_1=\SI{4}{\milli\electronvolt},\lambda_0=\SI{0}{\milli\electronvolt},\lambda_1=\SI{0}{\milli\electronvolt}$, (c) $t_1=\SI{4}{\milli\electronvolt},\lambda_0=\SI{1}{\milli\electronvolt},\lambda_1=\SI{0}{\milli\electronvolt}$, (d) $t_1=\SI{2}{\milli\electronvolt},\lambda_0=\SI{1}{\milli\electronvolt},\lambda_1=\SI{0.1}{\milli\electronvolt}$.}\label{fig:App_B}
	\end{figure}  
	
	The contribution from the vibrationally-assisted spin-orbit interaction to the electronic force must be considered separately, which we will not address here. Nonetheless, the population difference $\delta p$  successfully captures the voltage dependence of the dynamics of the average displacement $\bar{X}$.
	This analysis is justified by the parameter regime considered here, where the Peierls coupling dominates over the spin-orbit-vibrational coupling ($t_1> \lambda_1$). 
	
	\section{Voltage-Dependence of Spin-Selectivity Dynamics: Variation of Parameters}\label{app:B}
	
	To get a better understanding of the effect of the SOC and the electronic-vibrational coupling, in \cref{fig:App_B} the voltage dependent spin selectivity dynamics is presented for certain cases for the system with size $N=1\times 5$.   
	In particular, we consider the effect of SOC and vibrationally-assisted SOC in \cref{fig:App_B} (b) and (c), and the effect of the spin-independent electronic-vibrational coupling in \cref{fig:App_B} (d). To compare results, the spin selectivity for the original parameters is shown in \cref{fig:App_B} (a). 
	For the case of an achiral molecule, that is $\lambda_0=\lambda_1=0$ (\cref{fig:App_B} (b)), we observe an absence of spin selectivity for all voltages.  
	If the vibrationally-assisted SOC is zero, but the static SOC is non-zero (\cref{fig:App_B} (c)), the spin selectivity is slightly reduced in comparison to the model with original parameters. 
	Even so surprisingly, in the model with a slightly smaller spin-independent electronic-vibrational coupling (\cref{fig:App_B} (d)), spin selectivity is of the order of $S\sim 0.01\%$.
	
	\section{Convergence of the Spin Selectivity with Number of Trajectories}\label{app:C}
\begin{figure}[!h]
		\centering
		\includegraphics[width=0.99\columnwidth]{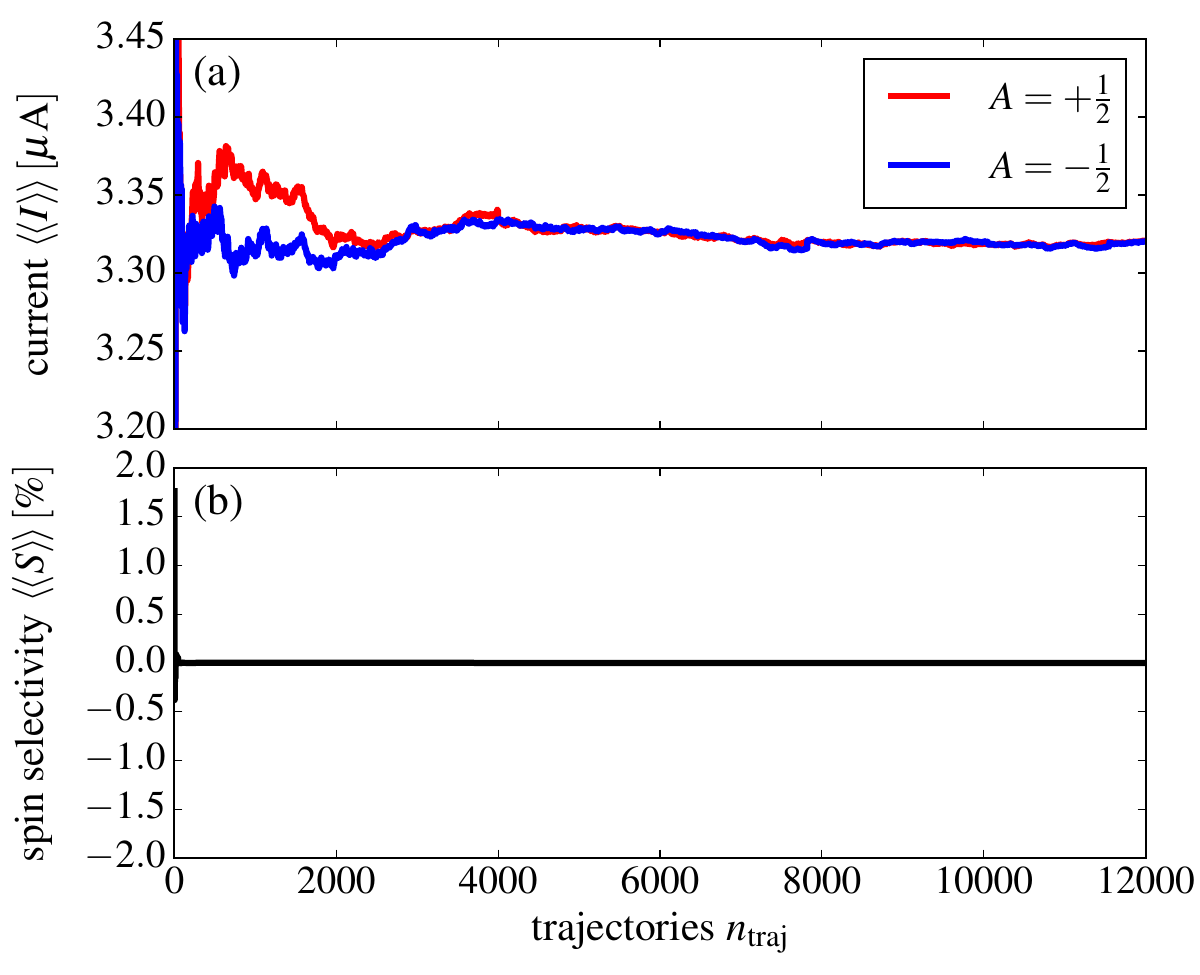}
		\caption{Dependence of time-averaged and ensemble averaged charge currents (a) and spin selectivity obtained from these currents (b) on the number of trajectories used in the ensemble average. Parameters are as in Sec.~\ref{sec:Results_C}.}\label{fig:App_C}
	\end{figure}

	As mentioned in the main text Sec.~\ref{sec:RC_2}, \cref{fig:App_C} displays the dependence of the time and ensemble averaged charge currents $\langle\!\langle I\rangle\!\rangle$ and the spin selectivity $\langle\!\langle S\rangle\!\rangle$ on the number of trajectories. 
	For an observable $O$, we define the time and ensemble average $\langle\!\langle O\rangle\!\rangle$ as
	\begin{equation}
		\langle\!\langle O\rangle\!\rangle=\frac{1}{T}\int_0^T \dif t \, \langle O(t)\rangle,\label{eq:Time_Ensemble_averaged_obs}
	\end{equation}
	with the ensemble-averaged observable $\langle O(t)\rangle$, defined in \cref{eq:Ensemble_averaged_obs}, and the integral in \cref{eq:Time_Ensemble_averaged_obs} is approximated by a Riemann sum with a time-step used in the numerical integrator.
	This result shows that time-averaged charge currents, evaluated in the time-interval $\SI{400}{\pico\second}$ until $\SI{800}{\pico\second}$. The former time is chosen such that from that time on the system is in a stationary state, and the latter is chosen for convenience to obtain sufficient temporal statistics maintaining a reasonable numerical cost.  
	In the main text, Sec.~\ref{sec:RC_2}, results for $n_\mathrm{traj}=12000$ trajectories are shown. From \cref{fig:App_C} (a), it is clear that the charge currents are well converged for this number of trajectories.
	With regard to the average steady-state spin selectivity, \cref{fig:App_C} (b) clearly shows that for this number of trajectories, time and ensemble-averaged spin selectivity is negligibly small.
	
	\bibliography{literature}
	
\end{document}